\documentclass[]{opendatalab}
\usepackage{xcolor}
\usepackage{longtable}
\usepackage{subcaption}
\usepackage{multirow}
\usepackage{graphicx}
\usepackage{hhline}
\usepackage{threeparttable}
\usepackage[table]{xcolor} 
\usepackage{booktabs}      
\definecolor{lightgray}{gray}{0.9} 
\usepackage{listings}
\usepackage[numbers]{natbib}
\definecolor{codegreen}{rgb}{0,0.6,0}
\definecolor{codegray}{rgb}{0.5,0.5,0.5}
\definecolor{codepurple}{rgb}{0.58,0,0.82}
\definecolor{backcolour}{rgb}{0.95,0.95,0.92}
\definecolor{promptcolor}{HTML}{D1D0F2}
\definecolor{promptcolorheader}{HTML}{bdbcec}
\usepackage{pifont}

\newcommand{\promptbox}[2]{
\begin{tcolorbox}[
top=0.3em,bottom=0.3em,left=0.5em,right=0.5em,
toptitle=0.3em,bottomtitle=0.2em,boxsep=0pt,
colframe=promptcolorheader,colback=promptcolor!50,boxrule=0.5pt,
]
\footnotesize
\end{tcolorbox}
}
\lstdefinestyle{mystyle}{
    backgroundcolor=\color{backcolour},   
    commentstyle=\color{codegreen},
    keywordstyle=\color{magenta},
    numberstyle=\tiny\color{codegray},
    stringstyle=\color{codepurple},
    basicstyle=\ttfamily\footnotesize,
    breakatwhitespace=false,         
    breaklines=true,                 
    captionpos=b,                    
    keepspaces=true,                 
    numbers=left,                    
    numbersep=5pt,                  
    showspaces=false,                
    showstringspaces=false,
    showtabs=false,                  
    tabsize=2
}

\title{Co-Scraper: query-aware DOM Pruning and Reusable Scraper Synthesis for Lightweight Web Data Extraction}

\author[1]{Shoupeng Wang}
\author[1]{Jiantao Qiu}
\author[2]{Wuyang Zhang}
\author[1]{Conghui He}

\affiliation[1]{Shanghai Artificial Intelligence Laboratory, OpenDataLab}
\affiliation[2]{University of Science and Technology of China}

\abstract{
The abundant and heterogeneous nature of web content necessitates automated information extraction, and generating scrapers that can be reused across similar web pages offers an effective solution for scalable data extraction. In this work, we propose Co-Scraper, a two-stage framework capable of handling the hierarchical complexity of long HTML documents. By integrating a query-aware DOM pruning mechanism with stable extraction strategy induction, Co-Scraper can effectively transforms web content into executable programmatic wrappers using a fine-tuned Qwen3-8B model. On the test set of SWDE, Co-Scraper achieves state-of-the-art performance with an F1 score of \textbf{94.78\%} and a reuse success rate of \textbf{90.39\%}.
This framework significantly enhances the accuracy and resilience of data extraction, providing a highly efficient approach for web data acquisition tasks.
}

\date{\today}
\correspondence{Jiantao Qiu, \email{qiujiantao@pjlab.org.cn}, Wuyang Zhang, \email{wuyangz@ustc.edu.cn}, Conghui He, \email{heconghui@pjlab.org.cn}}

\metadata[Code]{\url{https://github.com/Ambition-0303/Co-Scraper}}

\begin{document}

\maketitle

\section{Introduction}
\label{section:intro}

The web continuously produces a vast repository of information, and its industrial value lies in supporting scalable knowledge-base construction, price monitoring, and structured search, etc. 
However, the sheer volume and complexity of modern web content present a formidable challenge for efficient web data extraction (WDE).
Consequently, the development of agentic data acquisition frameworks with low cost and latency is essential for effective large-scale information retrieval. 
Using powerful large language models (LLMs) to process HTML documents with complex hierarchical structures and generate reusable scrapers has emerged as an effective paradigm for meeting this demand.

Traditional wrapper-based techniques require substantial manual effort to design and maintain extraction rules, which intrinsically constrains their scalability. Deep learning methods that model DOM structures reduce this reliance on manual rules, but they still fall short of the accuracy and generalization required for large-scale data extraction. LLM-driven methods further improve flexibility, yet they do not fully resolve the intertwined challenges of high cost, high latency, and low reusability. AutoScraper\cite{autoscraper} is a representative approach that uses LLMs to synthesize reusable scrapers, but it is limited to single-field extraction and still depends on closed-source LLMs.

A central cause behind these challenges is the hierarchical complexity of long HTML documents. In SWDE~\cite{swde}, the longest raw HTML reaches 219K tokens, and the website with the longest average raw HTML length reaches 91K tokens. Such inputs are disastrous not only because they may exceed practical context-window budgets, but also because their low signal-to-noise ratio substantially degrades model reasoning and extraction performance. This suggests that learning to process the DOM first is a more natural strategy than directly generating XPath rules over the full HTML. Studies such as HtmlRAG~\cite{htmlrag}, AXE~\cite{axe}, and Prune4Web~\cite{prune4web} similarly support that long-HTML noise must be explicitly compressed or filtered before being consumed by LLM-based systems.

\begin{figure}[htbp]
    \centering
    \includegraphics[width=0.95\textwidth]{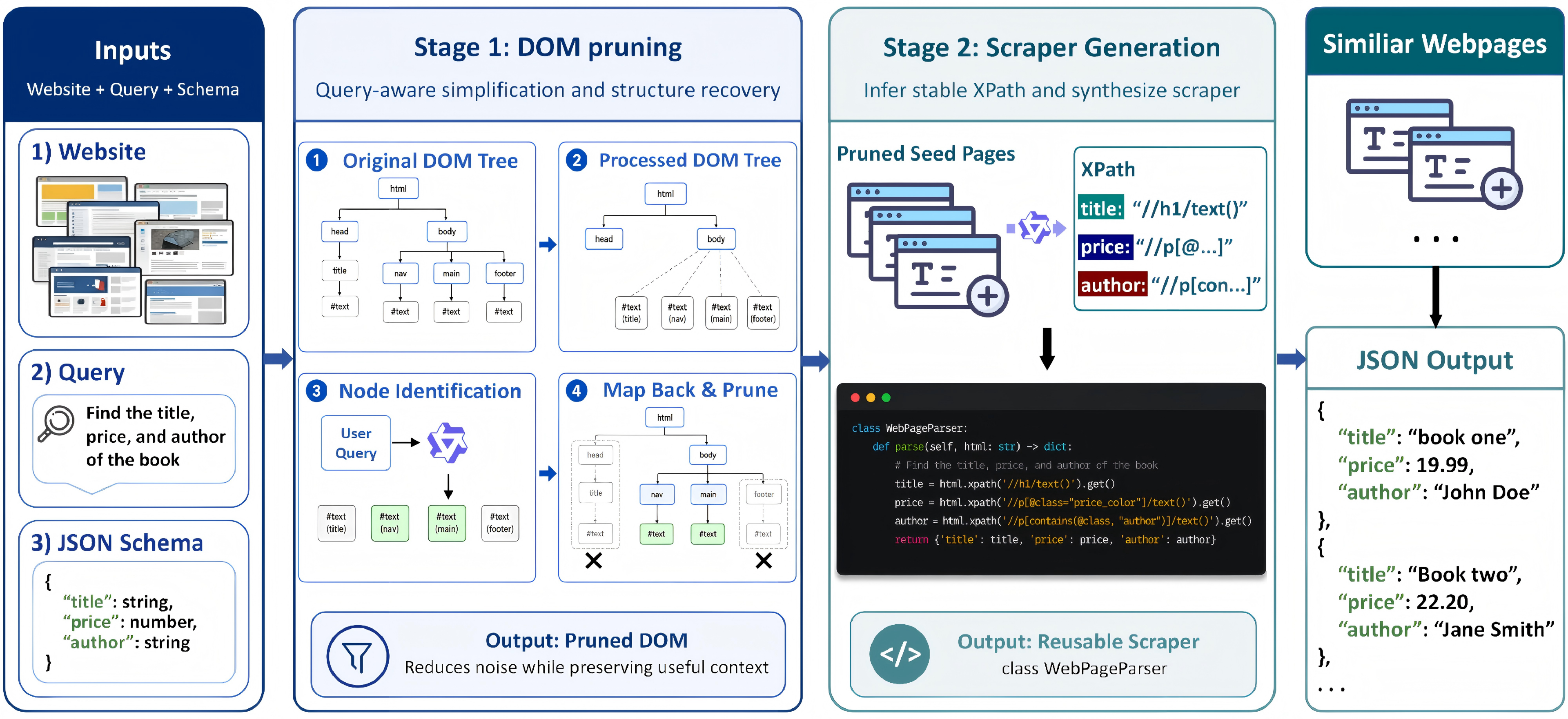}
    \caption{Two stage framework of Co-Scraper: DOM pruning and scraper generation.}
    \label{fig:coscraper}
\end{figure}

To address the aforementioned limitations, we propose Co-Scraper, a two-stage agentic framework
for robust, reusable web extraction using a small-scale language model. As illustrated in Figure~\ref{fig:coscraper},
Co-Scraper comprises two main components: DOM pruning and scraper generation. In the DOM pruning phase, the framework first converts the raw DOM into a simplified representation with a flatter and more manageable hierarchy, then identifies useful nodes that are relevant to the query. Finally, it performs pruning to remove redundant branches while preserving the essential document structure. This targeted compression yields a high-signal context that materially improves downstream reasoning quality.
In the scraper generation stage, Co-Scraper deduces the stable XPath corresponding to the required fields based on three pruned seed pages and synthesizes a programmatic wrapper which can be directly reused on similar web pages. Meanwhile, both stages use the same fine-tuned Qwen3-8B~\cite{qwen3} model named \textbf{Qwen-HTML}, which significantly reduces the inference cost and latency.
The primary contributions of this work are summarized as follows:
\begin{enumerate}
    \item \textbf{Two-Stage Agentic Framework for Scalable WDE: }We propose Co-Scraper, a two-stage agentic framework that combines query-aware DOM pruning with scraper generation, enabling effective large-scale web data acquisition.
    \item \textbf{Superior Accuracy and Reusability: }Co-Scraper supports simultaneous multi-field extraction and outperforms traditional methods and contemporary agentic baselines in both extraction accuracy and cross-page reusability, establishing a new state-of-the-art for autonomous WDE.
    \item \textbf{Edge-Ready Qwen-HTML: }By leveraging a small-scale model Qwen-HTML, Co-Scraper eliminates reliance on expensive APIs, significantly reducing both inference latency and operational expenditures while enhancing data privacy.
\end{enumerate}

\section{Related work}
\label{rew}

Traditional wrapper-based approaches, including WIEN~\cite{wien}, STALKER~\cite{stalker}, RoadRunner~\cite{roadrunner}, IEPAD~\cite{iepad}, and EXALG~\cite{exalg}, are effective under relatively regular page templates but remain brittle under template evolution and cross-site heterogeneity. Since these systems induce extraction rules from site-specific structural patterns, even small changes in the underlying HTML structure, such as inserting new DOM elements, modifying tag paths, or renaming class attributes, can invalidate previously learned wrappers. Maintaining such wrappers often requires manual inspection, wrapper re-induction, or rerunning the complete generation pipeline. This maintenance overhead becomes prohibitive in large-scale web extraction scenarios, limiting the applicability of classical wrapper-based techniques to today’s dynamic and heterogeneous web environment.

Automated WDE further requires to handle the inherently complex hierarchical structure of HTML DOM trees. With the development of deep learning methods, FreeDOM~\cite{freedom} learns transferable representations for DOM nodes, SimpDOM~\cite{simpdom} leverages a simplified tree representation to retrieve useful contexts, MarkupLM~\cite{markuplm} jointly pre‑trains text and markup signals for markup‑based documents, and WebFormer~\cite{webformer} designs HTML tokens with graph attention and builds rich attention patterns between HTML tokens and text tokens. However, these methods mainly formulate web extraction as node-level tagging or span prediction and typically rely on supervised training or task-specific fine-tuning. They remain sensitive to domain shift and large dynamically rendered DOMs, making it difficult to adapt real-world deeply nested structures and heterogeneous templates. 

In recent years, more effective DOM processing strategies combined with the powerful understanding capabilities of language models, have further improved the performance of WDE. Prune4Web~\cite{prune4web} reduces the burden of processing large and noisy DOM trees by shifting from direct LLM reading to programmatic DOM pruning, where LLM-generated executable scripts dynamically filter task-relevant elements before downstream grounding. AXE~\cite{axe} further demonstrates that lightweight extraction is possible by pruning boilerplate and irrelevant DOM nodes into a compact, high-density context, enabling a small language model to produce grounded structured outputs. However, it still performs inference-time extraction rather than producing reusable extraction logic across pages. AutoScraper~\cite{autoscraper} leverages the hierarchical structure of HTML and cross-page similarity to progressively generate executable action sequences. But it is confined to single‑field extraction and relies on frequent commercial API calls, significantly enhacing the inference cost and latency on large‑scale collection tasks. Therefore, existing methods still fall short of providing an efficient, lightweight, and truly scalable WDE that meets the demands of real‑world deployment.

\section{Methodology: Co-Scraper}
\label{method}

In this section, we describe the two-stage framework of Co-Scraper, including its query-aware DOM pruning mechanism, programmatic scraper generation, and the training procedure of Qwen-HTML.

\subsection{Task Formulation}
The purpose of our work is to achieve large-scale data extraction on a collection of similar web pages through a reusable scraper, rather than extracting values from a single page in isolation or extracting only a single field.
Let $\mathcal{H}$ denote a collection of similar webpages such as a website, $q$ denotes the user query containing the fields to be extracted,  and $s$ denotes the corresponding target schema. Co-Scraper is provided with three seed pages $\{h_1,h_2,h_3\}\subset\mathcal{H}_w$ from the same website. The goal is to synthesize a site-level extraction program $\pi_w$ such that, for any page $h\in\mathcal{H}_w$, the program produces a structured JSON:
\begin{equation}
    \hat{y}=\pi_w(h), \quad \forall h\in\mathcal{H}.
\end{equation}
The framework is explicitly decomposed into two stages, and both stages are powered by Qwen-HTML $\phi$. First, the query-aware DOM pruning module $P_\phi$ compresses each raw HTML page into an information-dense representation:
\begin{equation}
    P_\phi(h, q) \rightarrow \tilde{h}.
\end{equation}
At inference time, this pruning stage is conditioned only on the user query and the input HTML; it does not observe any gold attribute values. Second, the scraper generation module $G_\phi$ takes the three pruned seed pages and the schema as input, and generates the reusable site-level program:
\begin{equation}
    G_\phi(\tilde{h}_1,\tilde{h}_2,\tilde{h}_3,s) \rightarrow \pi_w.
\end{equation}

\subsection{DOM Pruning}
Raw HTML documents typically exhibit excessive length and low content density, characterized by redundant scripts and stylistic metadata. Directly processing such voluminous structures not only imposes a significant inference burden on LLMs but also adversely affects the accuracy of the output. Therefore, Co-Scraper first implements a query-aware HTML distillation to achieve an information-dense and structure-preserving representation.
\paragraph{Pre-processing}
The initial step removes elements that are irrelevant to the content such as \texttt{<head>}, \texttt{<style>}, and \texttt{<script>}, as well as invisible tags marked with properties like \texttt{display:none} and \texttt{font-size:0px}.
Due to the special nature where HTML text paragraphs and tag nodes may coexist under a single element, it is unsuitable to directly perform node identification. Therefore, we segment the HTML into distinct content blocks.
To ensure the atomicity of the blocks, we distinguish whether cohesive units such as lists (\texttt{<ul>}) and tables (\texttt{<table>}) serve layout or content purposes.
These blocks are then re-anchored as immediate descendants of the \texttt{<body>} element while preserving the original content order. To limit the length of each content block, we clean element attributes by retaining only valid \texttt{src}, \texttt{alt} attributes of image elements, and the \texttt{class}, \texttt{id} attributes of all elements. And textual paragraphs are truncated to their first 500 characters. For tables with numerous cells and lists with extensive items, we retain only representative subsets from the beginning and the end. At the same time, we record the mapping from the blocks to the original nodes for final distillation.
In this way, the input HTML is divided into parallel semantic blocks, generating two synchronized representations: simplified HTML, which removes the tags related to rendering to create a compact input for the LLM; and mapping HTML, which retains the original structure with complete element attributes and text segments to enable accurate reconstruction.

\paragraph{Node Identification}
After pre-processing, each semantic block is tagged with a unique identifier \texttt{\_item\_id}. Co-Scraper utilizes Qwen-HTML to process the simplified HTML with a parallel structure. Based on the captured query, the model will conduct semantic evaluation for each block and output a \texttt{useful\_item\_ids} list. This list contains all the items related to the target data, or the necessary elements as key structural anchors. By using a lightweight model, Co-Scraper maintains high precision in node identification while achieving extremely low inference latency, laying a highly efficient foundation for subsequent automated extraction.

\paragraph{Mapping Back}
In the final step, the system utilizes the identified \texttt{useful\_item\_ids} to reconstruct an information-dense representation from the mapping HTML. Co-Scraper adopts different processing methods for nodes at different positions. Specifically, for useful nodes, the framework preserves the complete information of the nodes themselves, their siblings, and the entire ancestral sequences to maintain structural and contextual integrity. For other useless nodes, the system preserves their lowest common ancestors, clearing the internal content while retaining their tags. Finally, the distilled HTML preserves the precise content and hierarchical structure of the nodes relevant to the query while significantly compressing the overall document length.

\subsection{Scraper Generation}
Following the DOM pruning, Qwen-HTML integrates the three pruned seed pages, the user
query and the JSON schema to synthesize deterministic and executable scrapers. By leveraging its enhanced HTML comprehension and stable XPath design capabilities, Co-Scraper ensures the accuracy and reusability of the extraction process across diverse web environments. Qwen-HTML is prompted to identify the required fields in three seed pages, carefully account for inter-page variations and design stable XPaths based on the priority strategy. The extraction strategy first prioritizes node-unique attributes and semantically meaningful descriptive class names, and then falls back to relative paths anchored under stable parent nodes and text matching. These XPaths are ultimately synthesized into a programmatic scraper that takes an HTML string as input and outputs a JSON dictionary.Compared with AutoScraper~\cite{autoscraper}, which directly uses XPaths as scrapers, the programmatic scraper provides more robust fallback and error-handling mechanisms, thereby achieving stronger stability and reusability.

\subsection{Qwen-HTML}
\subsubsection{Dataset}
\paragraph{SWDE~\cite{swde}} consists of 124,291 web pages spanning 8 distinct verticals: auto, book, camera, job, movie, nbaplayer, restaurant, and university. Each vertical has 3 to 5 fields to be extracted, and contains 10 different websites with each site providing between 200 and 2,000 individual pages. 

To avoid the distribution bias caused by class imbalance, we select at most 1,000 pages from each website to form the training and test set. Within the first seven verticals, we select eight websites from each domain for the training set, while the remaining two websites and the entirety of the \textit{university} domain are reserved for testing. As shown in Table~\ref{tab:swde}, this partitioning yields a training set of about 47,000 pages and a test set of about 21,000 pages. For DOM pruning, we utilize DeepSeek-V4-Pro~\cite{deepseekv4} to identify \texttt{useful\_item\_ids} and then filter the results by matching the characters with the ground truth. For scraper generation, we utilize Qwen3.5-397B-A17B~\cite{qwen35} to synthesize programmatic scrapers and filter them based on the execution results. Since each training instance of scraper generation requires three seed pages, we sample about 47,000 instances from the SWDE training split, with 40,000 used for supervised fine-tuning (SFT) and 7,000 used for reinforcement fine-tuning (RFT). The two training sets mentioned above are respectively named \textbf{SWDE-DOM} and \textbf{SWDE-scraper(SFT/RFT)}.

\subsubsection{Training}
\paragraph{SFT} Based on Qwen3-8B~\cite{qwen3}, we first conduct full-parameter fine-tuning on the SWDE-DOM and SWDE-scraper(SFT) datasets. Detailed training configurations are provided in Table~\ref{tab:sft_training_config}.
 This stage equips the model with fundamental HTML understanding, node identification, XPath design, and programmatic scraper generation capabilities.

\paragraph{RFT} After SFT, we further train Qwen-HTML with Group Relative Policy Optimization~\cite{deepseek-math} (GRPO) using SWDE-scraper(RFT). Detailed training configurations are provided in Table~\ref{tab:rft_training_config}. For each input $x$, GRPO samples a group of $G$ responses $\{y_i\}_{i=1}^{G}$ and optimizes the policy $\pi_{old}$ to the new policy $\pi_{\theta}$ with the group-normalized advantage:
\begin{equation}
\begin{aligned}
\mathcal{J}_{\mathrm{GRPO}}(\theta)
= &\frac{1}{G}\sum_{i=1}^{G}\frac{1}{|y_i|}\sum_{t=1}^{|y_i|}
\min\big(\rho_{i,t}(\theta)\hat{A}_i,
\mathrm{clip}(\rho_{i,t}(\theta),1-\epsilon,1+\epsilon)\hat{A}_i\big)
- \beta D_{\mathrm{KL}}(\pi_{\theta}\|\pi_{\mathrm{ref}}),
\end{aligned}
\end{equation}
where
\begin{equation*}
\rho_{i,t}(\theta)=\frac{\pi_{\theta}(y_{i,t}|x,y_{i,<t})}{\pi_{\theta_{\mathrm{old}}}(y_{i,t}|x,y_{i,<t})},\qquad \hat{A}_i=\frac{R_i-\mathrm{mean}(\{R_k\}_{k=1}^{G})}{\mathrm{std}(\{R_k\}_{k=1}^{G})+\delta}.
\end{equation*}
Here, $\epsilon$ controls the clipping range of the policy ratio, and $\beta \mathrel{>} 0$ controls the KL penalty with respect to the reference policy.
The reward function consists of two components. The first one is a format reward, which checks whether the response follows the required format with a valid \texttt{<think>} block and a parsable scraper code output:
\begin{equation}
R_{\mathrm{fmt}}(y)=\mathbf{1}\left[y\in\mathcal{F}\right],
\end{equation}
where $\mathcal{F}$ denotes the set of responses satisfying the required \texttt{<think>} and code-output format. The second one is a correctness reward, which executes the generated scraper $\pi_y$ on the three seed pages and evaluates the field-level F1 score:
\begin{equation}
R_{\mathrm{cor}}(y)=\frac{1}{3}\sum_{j=1}^{3}\frac{1}{|\mathcal{S}|}\sum_{f\in\mathcal{S}}\mathrm{F1}\left(\pi_y(h_j)_f, a_{j,f}\right),
\end{equation}
where $\mathcal{S}$ denotes the set of target fields, and $a_{j,f}$ is the ground-truth value of field $f$ on the seed page $h_j$. The final reward is the sum of the two components:
\begin{equation}
R(y)=R_{\mathrm{fmt}}(y)+R_{\mathrm{cor}}(y).
\end{equation}

\section{Experiments}
\label{exp}
In order to verify the effectiveness of Co-Scraper, this section mainly conducts experiments in the following aspects: 
(1) The accuracy of extraction on seed webpages.
(2) The reusability of scrapers on similar webpages.
(3) The execution efficiency of Co-Scraper specifically focusing on the inference latency.

\subsection{Experimental Settings \& Evaluation Metrics}

\paragraph{Experimental Settings} We evaluate Co-Scraper on the test set of SWDE with about 21,000 pages. These pages include websites that are excluded from the verticals covered by the training data, and websites form the truly out-of-domain vertical \textit{university}. All experimental results will be reported separately on university and non-University subsets to demonstrate the real generalization capability of our framework.

\paragraph{Evaluation Metrics} The primary workflow of Co-Scraper involves inferring a programmatic scraper from three seed pages and subsequently reusing the scraper across other pages. To rigorously evaluate the entire process, we assess the framework from three complementary perspectives: extraction accuracy, scraper reusability, and efficiency.

For extraction accuracy, we assess the two stages of the pipeline respectively. In the DOM pruning stage, we compare the predicted node set with the ground-truth node set and report set-level Precision, Recall, and F1 to measure whether task-relevant HTML regions are correctly preserved.  In the scraper generation stage, we prune the DOM tree based on ground-truth to generate the scraper. Then, we execute the generated scraper in a Docker environment to obtain deterministic extraction outputs, and evaluate each target field. Specifically, we report field-level Precision, Recall, and F1.  

Scraper reusability is evaluated from both field-level and page-level. 
Let $N_{\mathrm{total}}^{f}$ denote the total number of evaluated fields over all pages, $N_{\mathrm{success}}^{f}$ denote the number of fields whose precision, recall, and F1 are all equal to 1, and $N_{\mathrm{unex}}^{f}$ denote the number of fields with zero recall. The corresponding field-level metrics are defined as
\[
\mathrm{Cor}_{f} = \frac{N_{\mathrm{success}}^{f}}{N_{\mathrm{total}}^{f}},\qquad
\mathrm{Unex.}_{f} = \frac{N_{\mathrm{unex}}^{f}}{N_{\mathrm{total}}^{f}}.
\]
For page-level metrics, let $N_{\mathrm{total}}^{p}$ denote the total number of evaluated pages, $N_{\mathrm{success}}^{p}$ denote the number of pages on which all target fields are perfectly extracted with both precision, recall and F1 equal 1. The page-level metrics are formulated as
\[
\mathrm{Cor}_{p} = \frac{N_{\mathrm{success}}^{p}}{N_{\mathrm{total}}^{p}}.
\]

Finally, the two stages are combined using Qwen-HTML for both node identification and scraper generation to evaluate end-to-end extraction accuracy. And we take Gemini-3.5-Flash and other representative methods as baselines to compare with Co-Scraper. Meanwhile, efficiency is further assessed by reporting the latency of the full workflow. By comparing these runtimes against baseline methods, we verify the practical effectiveness and real-world viability of Co-Scraper. Together, these metrics provide a holistic view of the framework’s structural correctness, execution accuracy, cross-page generalization, and runtime cost.

\subsection{Results}
\paragraph{Extraction Accuracy \& Reusability}

\begin{table}[t]
    \centering
    
    \label{tab:combined_performance}
    
    \begin{subtable}{0.48\textwidth}
        \centering
        \renewcommand{\arraystretch}{1.2}
        \resizebox{\textwidth}{!}{ 
        \setlength{\tabcolsep}{3pt}
        \begin{tabular}{llccc}
            \toprule
            \textbf{Model} & \textbf{Category} & \textbf{Prec} & \textbf{Reca} & \textbf{F1} \\ \midrule
            \multirow{2}{*}{\textbf{DeepSeek-v4-Pro}} & Univ.     & 69.56 & 96.82 & 79.26 \\
                                           & Non-Univ. & 56.76 & 96.14 & 68.43 \\ \\[-4pt]
            \multirow{2}{*}{\textbf{Qwen-HTML}}      & Univ.     & 80.21 & 96.74 & 86.50 \\
                                           & Non-Univ. & 61.24 & 95.53 & 71.96 \\ \\[-4pt]
            \multirow{2}{*}{\textbf{Qwen3-8B}}      & Univ.     & 90.56  & 90.73  & 90.15 \\
                                           & Non-Univ. & 67.74 & 81.16 & 72.04  \\ \bottomrule
        \end{tabular}
        }
        \caption{DOM pruning performance}
        \label{tab:combined_performance_a}
    \end{subtable}
    \begin{subtable}{0.51\textwidth}
        \centering        
        \renewcommand{\arraystretch}{1.2}
        \resizebox{\textwidth}{!}{
        \setlength{\tabcolsep}{3pt} 
        \begin{tabular}{llccc}
            \toprule
            \textbf{Model} & \textbf{Category} & \textbf{Prec} & \textbf{Reca} & \textbf{F1} \\
            \midrule
            \multirow{2}{*}{\textbf{Qwen3.5-397B-A17B}}
            & Univ.       & 94.99 & 94.99 & 94.99  \\
            & Non-Univ.   & 88.31 & 88.57 & 88.39  \\
            \\[-4pt]
            \multirow{2}{*}{\textbf{Qwen-HTML}}
            & Univ.       & 96.98 & 96.98 & 96.98  \\
            & Non-Univ.   & 95.12 & 96.84 & 95.69  \\
            \\[-4pt]
            \multirow{2}{*}{\textbf{Qwen3-8B}}
            & Univ.     & 67.67 & 67.67 & 67.67 \\
            & Non-Univ. & 54.30 & 55.04 & 54.59  \\
            \bottomrule
        \end{tabular}
        }
        \caption{Scraper generation performance}
        \label{tab:combined_performance_b}
    \end{subtable}
    \vspace{0.6em}
    
    \begin{subtable}{\textwidth}
        \centering
        \renewcommand{\arraystretch}{1.2}
        \resizebox{\textwidth}{!}{
        \begin{tabular}{llcccccc}
            \toprule
             & \multirow{2}{*}{\textbf{Category}} & \multicolumn{3}{c}{\textbf{IE Evaluation}} & \multicolumn{3}{c}{\textbf{Executable Evaluation}} \\
             \cmidrule(lr){3-5} \cmidrule(lr){6-8}
             &  & Prec & Reca & F1 & Cor$_p$($\uparrow$) & Cor$_f$($\uparrow$) & Unex.$_f$($\downarrow$) \\
            \midrule
            \multirow{2}{*}{\textbf{Co-Scraper}}
            & Univ.       & 96.36 & 96.36 & 96.36 & 66.99 & 89.54 &  10.46 \\
            & Non-Univ.   & 94.39 & 95.58 & 94.78 & 70.73  & 90.39 & 7.34 \\
            \\[-4pt]
            \multirow{2}{*}{\textbf{Gemini-3.5-Flash}}
            & Univ.       & 99.55 & 99.55 & 99.55 & 69.20 & 89.68  & 10.32  \\
            & Non-Univ.   & 95.18 & 95.34 & 95.23 & 77.86 & 92.63  & 6.88  \\
            \\[-4pt]
            \textbf{AutoScraper}
            & -      & 92.49 & 89.13 & 88.69 & - & 71.56  & 4.06  \\
            \\
            \textbf{AXE}
            & -      & - & - & 88.10 & - & - & -   \\
            \bottomrule
        \end{tabular}
        }
        \caption{Co-Scraper performance with two stages combined}
        \label{tab:concat_performance}
    \end{subtable}
    \caption{Comprehensive performance evaluation of Co-Scraper on the test set of SWDE.}
\end{table}

The primary objective of node identification is to cover as many useful nodes rather than just the node where the answer lies, so we mainly focus on the recall of identification. Even if some extra nodes are included, the context length overhead is still much smaller than the original full HTML. Specifically, DOM pruning reduces the average HTML length in SWDE from 19.7K tokens to 0.7K tokens. Under this criterion, SFT clearly improves performance over the original 8B model as shown in Table~\ref{tab:combined_performance_a}. In the Non-Univ. split, recall increases from 81.13 to 95.53, which is close to DeepSeek-v4-Pro at 96.14. In the Univ. split, which is fully out-of-domain, recall also rises from 90.73 to 96.74, indicating that the trained model maintains generalization. For scraper generation, Table~\ref{tab:combined_performance_b} is evaluated using ground-truth node identification results to isolate scraper synthesis performance. The comparison shows that fine-tuning brings substantial gains over the base model Qwen3-8B with F1 rising from 67.67 to 96.98 on Univ. and from 54.59 to 95.69 on Non-Univ. More importantly, the RFT-enhanced Qwen-HTML also surpasses the much larger Qwen3.5-397B-A17B model, which achieves 94.99 F1 on Univ. and 88.39 F1 on Non-Univ.

Table~\ref{tab:concat_performance} evaluates the fully cascaded pipeline together with scrapper reusability under field-level and page-level executable metrics. After cascading the two trained components, Co-Scraper achieves stable end-to-end IE performance, reaching 96.36 F1 on Univ. and 94.78 F1 on Non-Univ.. From the executable perspective, Co-Scraper also maintains solid page-level reusability, with $\mathrm{Cor}_p=66.99$ on Univ. and $\mathrm{Cor}_p=70.73$ on Non-Univ. More importantly, the field-level results show that the generated scrappers remain highly usable even in the multi-field setting: Co-Scraper achieves $\mathrm{Cor}_f=89.54$ on Univ. and $\mathrm{Cor}_f=90.39$ on Non-Univ. And the performance of our fine-tuned Qwen-HTML is comparable to that of Gemini-3.5-Flash. Furthermore, Co-Scraper performs simultaneous multi-field extraction on each page, whereas AutoScraper only reports single-field extraction results. Under this stricter setting, Co-Scraper still achieves higher field-level reusability. This suggests that the proposed framework successfully extends scrapper-based extraction from single-field scenarios to realistic multi-attribute extraction while preserving strong robustness and cross-page reusability.

\paragraph{Efficiency}

\begin{table}[t]
\centering
\renewcommand{\arraystretch}{1.2}
\setlength{\tabcolsep}{10pt}
\begin{tabular}{lcccc}
\toprule
\textbf{Websites} & $T_i$ & $T_c$ & $T_{G_1}$ & $T_{G_2}$ \\
\midrule
Auto & 8.43s & 9.30s & 17.75s & 238.4s \\
Book & 5.66s & 9.94s & 15.63s & 176.4s \\
Camera & 8.64s & 8.24s & 16.91s & 107.1s \\
Job & 5.07s & 8.80s & 13.88s & 123.5s \\
Movie & 3.70s & 9.04s & 12.76s & 133.2s \\
Nbaplayer & 4.15s & 9.61s & 13.79s & 179.4s \\
Restaurant & 3.99s & 8.86s & 12.87s & 160.8s \\
University & 5.91s & 9.23s & 15.16s & 134.7s \\
\bottomrule
\end{tabular}
\caption{Latency of Co-Scraper across different websites.}
\label{tab:coscraper_latency}
\end{table}

We compare the overall extraction efficiency of Co-Scraper to another scraper-based method: AutoScraper. 
The latency of scraper-based methods can be uniformly formulated as $T = T_{G} + T_{E}$, where
$T_{G}$ denotes the generation time of the scraper and $T_{E}$ denotes the execution time of the scraper on other webpages. 
For Co-Scraper, let $T_i$ denote the node identification time, $T_p$ denote the pruning execution time, $T_s$ denote the scraper generation time, $N_W$ denote the number of webpages to be processed on the same website, and $T_e$ denote the execution time of the scraper on one webpage. Then the total time of our method is
\begin{equation}
T_1 = T_{G_1} + T_{E_1} = (T_i + T_p + T_s) + N_W T_e.
\end{equation}
For AutoScraper, suppose the number of seed webpages is $n_s$, the time to generate a wrapper is $T_g$, and the time for wrapper synthesis is $T_x$. The total time can be formulated as
\begin{equation}
T_2 = T_{G_2} + T_{E_2} =  (n_s T_g + T_x) + N_W T_e.
\end{equation}


The inference of Co-Scraper is accomplished on one H200 GPU using vLLM~\cite{vllm}, and each sample is executed sequentially.
Table~\ref{tab:coscraper_latency} shows that Co-Scraper achieves substantially lower wrapper-construction latency than AutoScraper across all website categories. Specifically, the total generation time of Co-Scraper ($T_{G_1}$) remains within 12.76s--17.75s, whereas the corresponding AutoScraper generation time ($T_{G_2}$) ranges from 107.1s to 238.4s. This large gap indicates that Co-Scraper produces reusable scrappers much more efficiently than AutoScraper. When $N_W$ is large, both scrapper-based paradigms are substantially more efficient than direct LLM extraction, and Co-Scraper further improves practical efficiency by reducing the generation cost of reusable scrappers.

\section{Discussion}
\subsection{Ablation Study}

To verify whether DOM pruning directly improves HTML processing performance in WDE, Table~\ref{tab:ablation} compares Qwen3-8B under three input settings: raw HTML, HTML pruned by the base model, and HTML pruned based on ground-truth nodes. Compared with direct inference on raw HTML, using ground-truth pruned HTML improves F1. It shows that removing noisy DOM content can directly enhance extraction capability of the base model. However, pruning with the untrained Qwen3-8B does not yield a reliable gain. These results indicate reliable DOM pruning is the foundation for the high performance of the Co-Scraper framework.

\subsection{Comparison with supervised learning models}
We conduct a comparison with 5 supervised learning baseline models in WDE: Render-Full~\cite{renderfull}, FreeDOM~\cite{freedom}, SimpDOM~\cite{simpdom}, MarkupLM$_{\textsc{BASE}}$~\cite{markuplm} and WebFormer~\cite{webformer}. The results are shown in Table~\ref{tab:baseline_results} and Co-Scraper uses the result of Univ. which is fully out-of-domain.

\begin{table}[t]
\centering

\begin{subtable}[t]{0.58\textwidth}

\centering

\resizebox{\linewidth}{!}{
\begin{tabular}{llccc}
\toprule
\textbf{Model} & \textbf{Category} & \textbf{Prec} & \textbf{Reca} & \textbf{F1} \\
\midrule
Pruned HTML$_{gt}$ & Univ.     & 67.67 & 67.67 & 67.67 \\
+Qwen3-8B & Non-Univ. & 54.30 & 55.04 & 54.59  \\
\\[-4pt]
Pruned HTML$_{base}$ & Univ.     & 52.58 & 52.58 & 52.58 \\
+Qwen3-8B & Non-Univ. & 53.59 & 53.57 & 53.57  \\
\\[-4pt]
Raw HTML & Univ.     & 58.04 & 58.04 & 58.04 \\
+Qwen3-8B & Non-Univ. & 48.34 & 48.29 & 48.31  \\
\bottomrule
\end{tabular}
}
\caption{Performance comparison of Qwen3-8B with different HTML inputs. Pruned HTML$_{gt}$ and Pruned HTML$_{base}$ respectively represent the pruning based on the ground truth and the pruning using the untrained Qwen3-8B model.}
\label{tab:ablation}
\end{subtable}%
\hfill
\begin{subtable}[t]{0.39\textwidth}
\centering

\begin{tabular*}{\linewidth}{@{\extracolsep{\fill}}lc}
\toprule
\textbf{Model} & \textbf{F1} \\
\midrule
Render-Full~\cite{renderfull}        & 84.30 \\[+2.6pt]
FreeDOM~\cite{freedom}               & 82.32 \\[+2.6pt]
SimpDOM~\cite{simpdom}               & 83.06 \\[+2.6pt]
MarkupLM$_{\textsc{BASE}}$~\cite{markuplm} & 84.31 \\[+2.6pt]
WebFormer~\cite{webformer}           & 86.58 \\
\midrule
Co-Scraper                  & \textbf{94.78} \\
\bottomrule
\end{tabular*}
\caption{Comparison of 5 supervised learning models to Co-Scraper on the SWDE dataset. Each value of the supervised model in the table is trained on 1 seed site.}
\label{tab:baseline_results}
\end{subtable}

\caption{\textbf{(a)} Ablation study of DOM pruning. \textbf{(b)} Comparison with supervised learning baselines.}
\label{tab:discussion}

\end{table}

\section{Conclusion}
\label{conclusion}
Compared with wrapper-based and supervised learning-based paradigms, scraper-based methods have inherent advantages in large-scale deployment and cross-page reusability, while still benefiting from strong language-model reasoning to achieve higher extraction accuracy. Within the scraper-based family, Co-Scraper further advances this direction through a better two-stage framework design. By coupling query-aware node identification with robust code generation, Co-Scraper consistently outperforms AutoScraper in accuracy, reusability, efficiency, and overall cost-effectiveness, demonstrating a practical and scalable solution for autonomous WDE.

\clearpage
\newpage
\bibliographystyle{unsrtnat}
\setcitestyle{numbers}
\bibliography{paper}

\clearpage
\newpage
\beginappendix

\section{Dataset Statistic}

\begin{table}[h]
\centering

\begin{minipage}[t]{0.48\textwidth}
\centering
\footnotesize
\begin{tabular}{lllc}
\toprule
\textbf{Vertical} & \textbf{Attributes} & \textbf{Website} & \textbf{Pages} \\
\midrule
\multirow{10}{*}{\textbf{Auto}} & \multirow{10}{*}{\begin{tabular}[c]{@{}l@{}} model \\ price \\ engine \\ fuel \end{tabular}} & aol & 1,000 \\
    & & autobyte & 1,000 \\
    & & automotive & 1,000 \\
    & & autoweb & 1,000 \\
    & & carquotes & 1,000 \\
    & & cars & 1,000 \\
    & & kbb & 1,000 \\
    & & motortrend & 1,000 \\ 
\rowcolor{lightgray} \cellcolor{white} & \cellcolor{white} & msn & 1,000 \\ 
\rowcolor{lightgray} \cellcolor{white} & \cellcolor{white} & yahoo & 1,000 \\ \midrule
\multirow{10}{*}{\textbf{Book}} & 
\multirow{10}{*}{\begin{tabular}[c]{@{}l@{}} title \\ author \\ isbn\_13 \\ publisher \\ publication\_date \end{tabular}} & 
    abebooks & 1,000 \\
    & & amazon & 1,000 \\
    & & barnesandnoble & 1,000 \\
    & & bookdepository & 1,000 \\
    & & booksamillion & 1,000 \\
    & & bookorders & 1,000 \\
    & & buy & 1,000 \\
    & & christianbook & 1,000 \\ 
\rowcolor{lightgray} \cellcolor{white} & \cellcolor{white} & deepdiscount & 1,000 \\ 
\rowcolor{lightgray} \cellcolor{white} & \cellcolor{white} & waterstone & 1,000 \\ \midrule
\multirow{10}{*}{\textbf{Auto}} & \multirow{10}{*}{\begin{tabular}[c]{@{}l@{}} model \\ price \\ manufacturer \end{tabular}} & amazon & 1,000 \\
    & & beachaudio & 247 \\
    & & buy & 500 \\
    & & compsource & 430 \\
    & & ecost & 923 \\
    & & jr & 367 \\
    & & newegg & 220 \\
    & & onsale & 261 \\ 
\rowcolor{lightgray} \cellcolor{white} & \cellcolor{white} & pcnation & 234 \\ 
\rowcolor{lightgray} \cellcolor{white} & \cellcolor{white} & thenerd & 309 \\ \midrule
\multirow{10}{*}{\textbf{Job}} & \multirow{10}{*}{\begin{tabular}[c]{@{}l@{}} title \\ company \\  location \\ date\_posted \end{tabular}} & careerbuilder & 1,000 \\
    & & dice & 1,000 \\
    & & hotjobs & 1,000 \\
    & & job & 1,000 \\
    & & jobcircle & 1,000 \\
    & & jobtarget & 1,000 \\
    & & monster & 1,000 \\
    & & nettemps & 1,000 \\ 
\rowcolor{lightgray} \cellcolor{white} & \cellcolor{white} & rightitjobs & 1,000 \\ 
\rowcolor{lightgray} \cellcolor{white} & \cellcolor{white} & techcentric & 1,000 \\
\bottomrule
\end{tabular}
\end{minipage}
\begin{minipage}[t]{0.48\textwidth}
\centering
\footnotesize
\begin{tabular}{lllc}
\toprule
\textbf{Vertical} & \textbf{Attributes} & \textbf{Website} & \textbf{Pages} \\
\midrule
\multirow{10}{*}{\textbf{Movie}} & \multirow{10}{*}{\begin{tabular}[c]{@{}l@{}} title \\ director \\ genre \\ mpaa\_rating \end{tabular}} & allmovie & 1,000 \\
    & & amctv & 1,000 \\
    & & boxofficemojo & 1,000 \\
    & & hollywood & 1,000 \\
    & & iheartmovies & 1,000 \\
    & & imdb & 1,000 \\
    & & metacritic & 1,000 \\
    & & msn & 1,000 \\ 
\rowcolor{lightgray} \cellcolor{white} & \cellcolor{white} & rottentomatoes & 1,000 \\ 
\rowcolor{lightgray} \cellcolor{white} & \cellcolor{white} & yahoo & 1,000 \\ \midrule
\multirow{10}{*}{\textbf{NBAPlayer}} & \multirow{10}{*}{\begin{tabular}[c]{@{}l@{}} name \\ team \\ height \\ weight \end{tabular}} & espn & 434 \\
    & & fanhouse & 446 \\
    & & foxsports & 425 \\
    & & msnca & 434 \\
    & & nba & 434 \\
    & & si & 515 \\
    & & slam & 423 \\
    & & usatoday & 436 \\ 
\rowcolor{lightgray} \cellcolor{white} & \cellcolor{white} & wiki & 420 \\ 
\rowcolor{lightgray} \cellcolor{white} & \cellcolor{white} & yahoo & 438 \\ \midrule
\multirow{10}{*}{\textbf{Restaurant}} & \multirow{10}{*}{\begin{tabular}[c]{@{}l@{}} name \\ address \\ phone \\ cuisine \end{tabular}} & fodors & 1,000 \\
    & & frommers & 1,000 \\
    & & zagat & 1,000 \\
    & & gayot & 1,000 \\
    & & opentable & 1,000 \\
    & & pickaretaurant & 1,000 \\
    & & restaurantica & 1,000 \\
    & & tripadvisor & 1,000 \\ 
\rowcolor{lightgray} \cellcolor{white} & \cellcolor{white} & urbanspoon & 1,000 \\ 
\rowcolor{lightgray} \cellcolor{white} & \cellcolor{white} & usdiners & 1,000 \\ \midrule
\multirow{10}{*}{\textbf{University}} & 
\multirow{10}{*}{\begin{tabular}[c]{@{}l@{}} name \\ phone \\ website \\ type \end{tabular}} & 
    \cellcolor{lightgray} collegeboard &\cellcolor{lightgray} 1,000 \\
    & &\cellcolor{lightgray} collegenavigator & \cellcolor{lightgray}1,000 \\
    & &\cellcolor{lightgray} collegeprowler &\cellcolor{lightgray} 1,000 \\
    & &\cellcolor{lightgray} collegetoolkit &\cellcolor{lightgray} 1,000 \\
    & & \cellcolor{lightgray}ecampustours &\cellcolor{lightgray} 1,000 \\
    & &\cellcolor{lightgray} embark &\cellcolor{lightgray} 1,000 \\
    & & \cellcolor{lightgray}matchcollege & \cellcolor{lightgray}1,000 \\
    & & \cellcolor{lightgray}princetonreview & \cellcolor{lightgray}615 \\ 
    & &\cellcolor{lightgray} studentaid &\cellcolor{lightgray} 1,000 \\ 
    & &\cellcolor{lightgray} usnews &\cellcolor{lightgray} 1,000 \\
\bottomrule
\end{tabular}
\end{minipage}
\caption{Detailed website-level statistics of SWDE: The total number  of the training and test set are 47,061 and 21,016 respectively. Test sites are marked in \colorbox{lightgray}{gray}.}
\label{tab:swde}
\end{table}

\section{Training Configure}
\label{sec:training-configure}

Table~\ref{tab:training_config} shows the specific training parameters of Qwen-HTML. Both SFT and RFT are conducted on 8$\times$H200 GPUs.

\begin{table}[h]
\centering
\small
\begin{subtable}[t]{0.48\textwidth}
\centering
\begin{tabular}{ll}
\hline
\textbf{Configuration} & \textbf{Value} \\
\hline
Base model & Qwen3-8B \\
DeepSpeed strategy & ZeRO-3 \\
Training datasets & SWDE-DOM, SWDE-scraper(SFT) \\
Batch size & 32 \\
Learning rate & $1.0 \times 10^{-5}$ \\
Epochs & 3 \\
Warmup ratio & 0.1 \\
Precision & BF16 \\
Fine-tuning type & Full-parameter \\
\hline
\end{tabular}
\caption{SFT}
\label{tab:sft_training_config}
\end{subtable}
\hfill
\begin{subtable}[t]{0.48\textwidth}
\centering
\begin{tabular}{ll}
\hline
\textbf{Configuration} & \textbf{Value} \\
\hline
Rollout number & 4 \\
DeepSpeed strategy & ZeRO-3 \\
Training datasets & SWDE-scraper(RFT) \\
Batch size & 16 \\
Learning rate & $1.0 \times 10^{-6}$ \\
Epochs & 3 \\
Warmup steps & 50 \\
Precision & BF16 \\
KL coefficient $\beta$ & 0.001 \\
\hline
\end{tabular}
\caption{RFT}
\label{tab:rft_training_config}
\end{subtable}
\caption{Training configuration of Qwen-HTML in SFT and RFT.}
\label{tab:training_config}
\end{table}

\section{Case Study}

\begin{figure}[h]
    \centering
    \includegraphics[width=\textwidth]{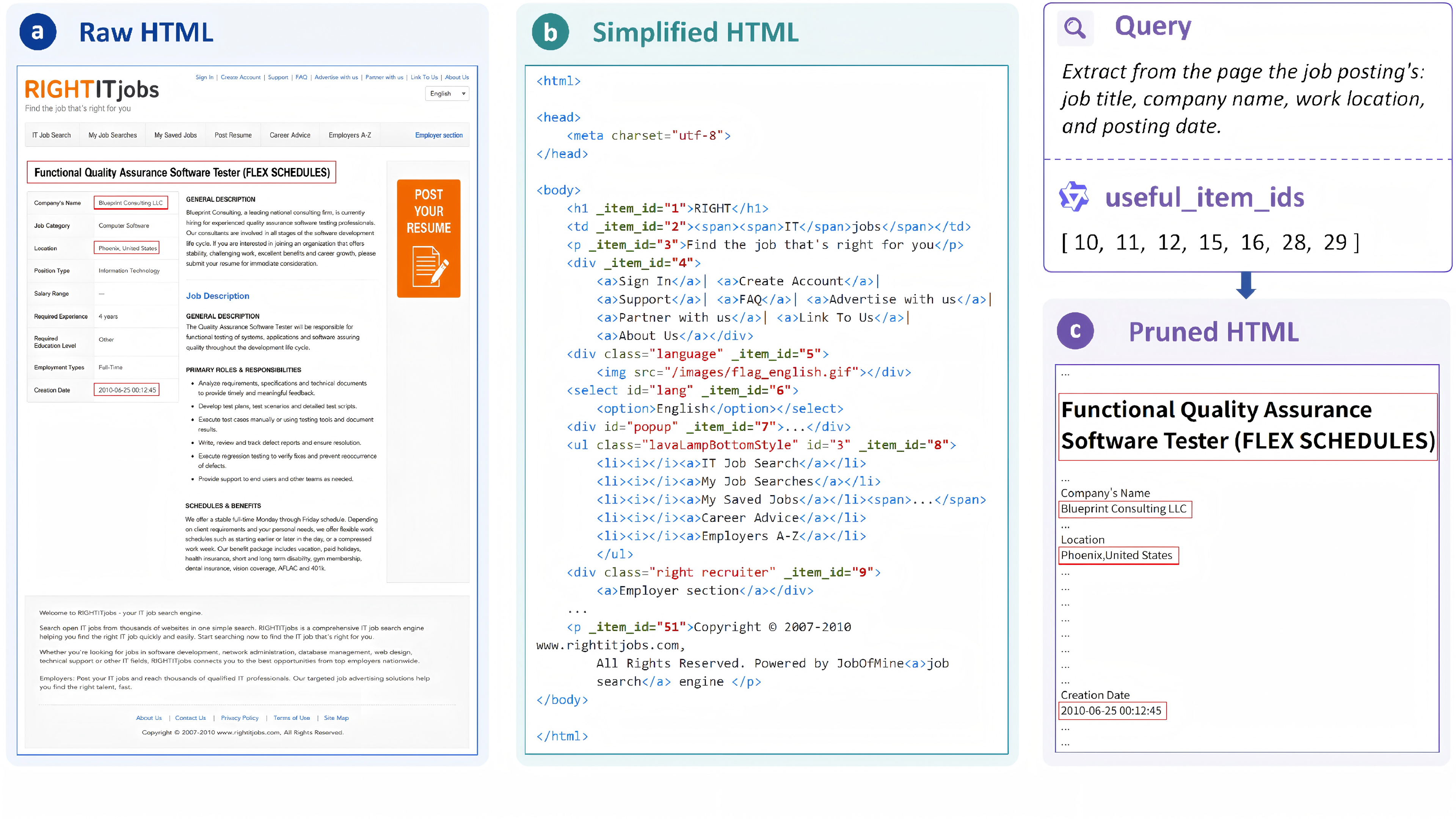}
    \caption{Case of DOM pruning: \textbf{(a)} raw HTML page of a job posting with a nested structure and a large amount of irrelevant content; \textbf{(b)} simplified HTML with all node adhered to \textit{<body>} tag; \textbf{(c)} pruned HTML retaining only task-relevant elements with original DOM structure preserved based on the identified useful item IDs. }
    \label{fig:cs1}
\end{figure}

\begin{figure}[t]
    \centering
    \includegraphics[width=\textwidth]{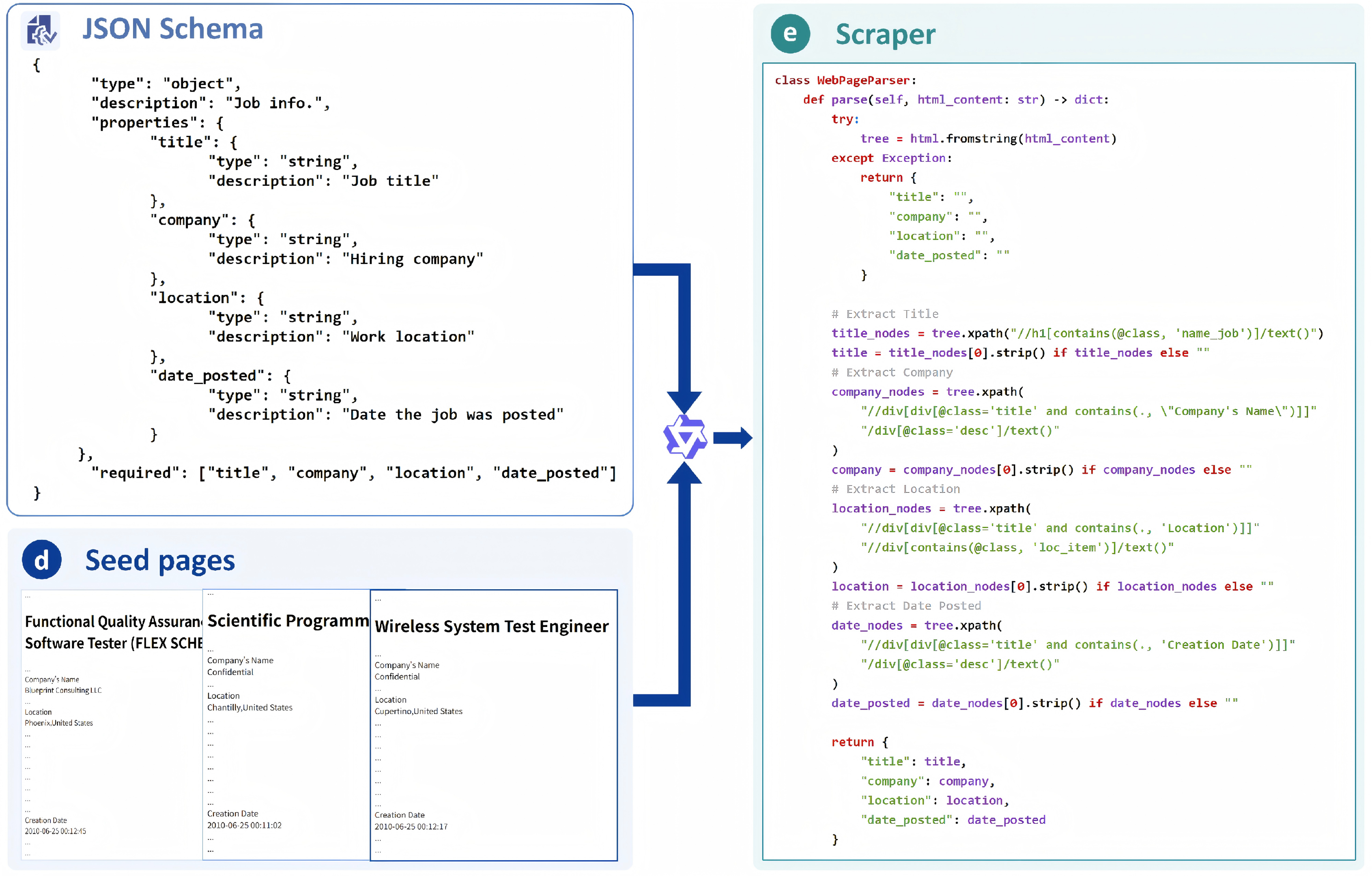}
    \caption{Case of scraper generation: \textbf{(d)} three pruned seed pages; \textbf{(e)} reusable scraper based on JSON schema and seed pages. }
    \label{fig:cs1}
\end{figure}

\end{document}